\documentclass[twocolumn,superscriptaddress,showpacs, aps, prb]{revtex4-2}
\usepackage{color}
\usepackage{hyperref}
\usepackage{amsmath, amssymb}
\hypersetup{colorlinks=true,breaklinks,linkcolor=blue,urlcolor=blue,citecolor=blue}
\usepackage{graphicx}% Include figure files
\usepackage{dcolumn}% Align table columns on decimal point
\usepackage{bm}% bold math
\usepackage{xspace}
\usepackage{braket}

\newcommand{\tiG}{\tilde{G}}
\newcommand{\tiR}{\tilde{\rho}}
\newcommand{\tiRvec}{\tilde{\bm{\rho}}}
\DeclareMathOperator*{\minarg}{\mathrm{minarg}}

\newcommand{\Pade}{Pad\'{e}\xspace}

\usepackage{natbib}
\begin{document}

\preprint{}

\title{
Robust analytic continuation combining the advantages of \\
the sparse modeling approach and Pad\'{e} approximation
%Robust analytic continuation by combining the sparse modeling method with the \Pade approximation method
}% Force line breaks with \\

\author{Yuichi Motoyama}
\affiliation{Institute for Solid State Physics, University of Tokyo, Chiba 277-8581, Japan}
\author{Kazuyoshi Yoshimi}
\affiliation{Institute for Solid State Physics, University of Tokyo, Chiba 277-8581, Japan}
\author{Junya Otsuki}
\affiliation{Research Institute for Interdisciplinary Science, Okayama University, Okayama 700-8530, Japan}

\date{\today}% It is always \today, today,

\begin{abstract}
Analytic continuation (AC) from the imaginary-time Green's function to the spectral function is a crucial process for numerical studies of the dynamical properties of quantum many-body systems.
This process, however, is an ill-posed problem; that is, the obtained spectrum is unstable against the noise of the Green's function.
Though several numerical methods have been developed, each of them has its own advantages and disadvantages.
The sparse modeling (SpM) AC method, for example, is robust against the noise of the Green's function 
%but suffers from unphysical oscillation in the obtained spectrum.
but suffers from unphysical oscillations in the low-energy region.
We propose a method that combines the SpM AC with the \Pade approximation.
This combination, called SpM-\Pade, inherits robustness against noise from SpM and low-energy accuracy from \Pade, compensating for the disadvantages of each.
We demonstrate that the SpM-\Pade method yields low-variance and low-biased results with almost the same computational cost as that of the SpM method.

\end{abstract}

\maketitle

%\tableofcontents

\section{Introduction}

The imaginary-time representation provides a foundation of calculations for quantum many-body systems at finite temperature, both analytically and numerically.
In analytical calculations, the thermal expectation values of several static and dynamical quantities can be calculated from the imaginary-time Green's function using the diagram technique~\cite{AbrikosovGD1963}.
In numerical calculations, path-integral Monte Carlo (PIMC) methods based on the imaginary-time representation offer ways to calculate the expectation values of several kinds of physical quantities at finite temperature~\cite{GullMLRTW2011, GubernatisKW2016}.
Besides classical quantities such as density and energy, recent developments enable the PIMC methods to directly calculate some topological ones of the wave function such as the susceptibility of the fidelity~\cite{SchwandtAC2009, WangLIMT2015}, the Berry curvature~\cite{Kolodrubetz2014}, and the Berry phase~\cite{MotoyamaT2013,MotoyamaT2018}. 
Dynamical quantities such as a spectrum function, however, cannot be calculated directly at this time, and we need to perform an analytic continuation (AC) of the Green's function from the imaginary time (frequency) to the real one.

The transform in AC is known as an ill-posed inverse problem; that is, the obtained spectrum is strongly affected by the noise or uncertainties of the imaginary-time Green's function, as seen later.
Since this noise is unavoidable in the PIMC method, several methods for reconstructing the spectrum have been developed in order to overcome this ill-posed problem over the years:
the \Pade approximation~\cite{VidbergS1977} and its variant~\cite{Kiss2019,Weh2020}, the maximum entropy method (MaxEnt)~\cite{SilverSG1990, JarrelG1996}, the regression using deep neural networks (DNN)~\cite{YoonSH2018, FournierWYW2020, KadesPRSUWWZ2020, XieBMW2021}, the stochastic AC method~\cite{Sandvik1998, MishchenkoPSS2000, Beach2004, FuchsPJ2010, Sandvik2016, ShaoQCCMS2017}, and the sparse modeling (SpM) method~\cite{OtsukiOSY2017, SpMReview}.
These methods each have their own advantages and disadvantages.
The \Pade approximation is simple and fast for calculation but is weak against noise.
The MaxEnt method is stable but requires knowledge or intuition about the spectrum function, the so-called default model.
DNN requires a large number of examples (pairs of Green's function and spectra) and a long time is required to train the regression DNN model.
Once the model is trained, a spectrum can be regenerated very rapidly.
The stochastic AC method is stable and requires no default model, but an extra Monte Carlo sampling of spectrum functions is needed.
The SpM method is a stable method not requiring any domain knowledge, but gives an artificial oscillation in the obtained spectrum~\footnote{Note that other methods using basis transformation and truncation also face similar oscillation. Oscillation in the SpM method, however, seems larger than that in others.}.

In this paper, we propose another AC method, the SpM-\Pade method, combining the SpM method with the \Pade approximation method.
This inherits strong points from the parent methods: robustness against the noise of the input from SpM and smoothness and correctness in the low-frequency region from \Pade.
As with the SpM method, additionally, the SpM-\Pade method requires no intuition about the system and no lengthy preprocess.

The structure of this paper is as follows:
We first briefly describe the problem to be solved in section \ref{sec:problem}.
In section \ref{sec:method}, we first review the \Pade approximation method and the SpM method briefly, and then introduce the proposed method, SpM-\Pade{}.
Next, we demonstrate the methods in section \ref{sec:demonstration}.
Finally, we summarize the paper in section \ref{sec:summary}.

\section{Problem to be solved}
\label{sec:problem}
The spectrum function $\rho(\omega)$ can be reconstructed from an imaginary-time Green's function, $G(\tau)$, via the following integral equation:
\begin{equation}
    G(\tau) = \int_{-\infty}^\infty d\omega K(\tau, \omega) \rho(\omega),
\label{eq:inteq}
\end{equation}
where $\tau$ is the imaginary time, and $\omega$ is the real frequency.
The integral kernel is 
\begin{equation}
    K(\tau, \omega) = \frac{e^{-\tau \omega}}{1 \pm e^{-\beta \omega}}, 
\end{equation}
where $\beta = 1/k_\text{B} T$ is the inverse temperature.
The plus or minus sign in the denominator depends on the statistics of the system:
plus for a fermionic system and minus for a bosonic one.
Since $K(\tau,\omega)$ becomes exponentially smaller as $\tau$ and $\omega$ become larger, a small change in $\rho(\omega)$ in the high-frequency region has exponentially small effects on $G(\tau)$.
A noise of $G(\tau)$, inversely, is magnified in $\rho(\omega)$.
As a result, the inverse problem (to obtain $\rho$ from given noisy $G$) is ill posed.

\section{Methods}
\label{sec:method}
First, we will briefly review two existing methods for performing AC, the \Pade approximation method and the SpM method.
Then, we will introduce the proposed method, SpM-\Pade{}, which combines the former two methods.

\subsection{\Pade approximation}
The first method, the \Pade approximation, is the simplest and the fastest method for estimating the spectrum function.
First, perform a Fourier transform of the imaginary-time Green's function $G(\tau)$ to the imaginary-frequency Green's function $G(i\omega_n)$, where $\omega_n$ are Matsubara frequencies.
Next, fit some function, say $\bar{G}(i\omega)$, to $G(i\omega_n)$ and replace $i\omega$ in $\bar{G}(i\omega)$ with $\omega + i\delta$, where $\delta > 0$ is a very small constant for the purpose of avoiding poles on the real axis.
Once the real-frequency Green's function $\bar{G}(\omega+i\delta)$ is obtained, the spectrum function can be calculated as $\rho(\omega) = -\mathrm{Im}\bar{G}(\omega+i\delta)/\pi$.
In this paper we adopt a continued fraction for $\bar{G}(i\omega)$ (the \Pade approximation) as follows:

%\color{red}
\begin{equation}
\bar{G}(\omega) = 
\cfrac{a_0}{1 + \cfrac{a_1(\omega-i\omega_1)}{1 + \cfrac{a_2(\omega - i\omega_2)}{1+\cdots}}},
\end{equation}
where $\{a_i\}$ are the fitting parameters.
%\color{black}

The \Pade method is easy to implement and fast to calculate.
As seen later, while the obtained spectrum seems smooth and accurate in the small-frequency region,
it becomes worse in the high-frequency region, especially beyond a peak.
This is because the direct AC scheme from $\bar{G}(i \omega)$ to $\bar{G}(\omega + i\delta)$ is an extrapolation from the imaginary axis to the real axis, and most of the information is used to construct the first peak.

\subsection{Sparse Modeling (SpM)}
The second method, SpM, is used to remove the noise by using the SpM method.
First, we discretize $\tau$ and $\omega$ in Eq.~(\ref{eq:inteq}) as follows:
\begin{equation}
    G_i = \sum_j K_{ij} \rho_j,
\label{eq:disceq}
\end{equation}
where $G_i = G(\tau_i)$ and $\rho_j = \rho(\omega_j)\Delta\omega$, and $\Delta\omega = (\omega_\text{max} - \omega_\text{min})/(N_\omega-1)$.
Next, we perform singular value decomposition of the kernel matrix as $K_{ij} = \sum_\ell U_{i\ell} S_\ell V_{\ell j}^t$ and transform the basis by using the obtained singular vectors as
\begin{equation}
    \tiG_\ell = \sum_i U_{\ell i}^t G_i
\end{equation}
and
\begin{equation}
    \tiR_\ell = \sum_j V_{\ell j}^t \rho_j.
\end{equation}
Keep in mind that a symbol with a tilde mark denotes a quantity represented in the new basis, called the intermediate representation (IR) basis~\cite{ShinaokaOOY2017}.
As a result, the following simple form,
\begin{equation}
    \tiG_\ell = S_\ell \tilde{\rho}_\ell
\end{equation}
is obtained.
It should be noted that when the singular values of the kernel $S_\ell$ decay exponentially, we can safely truncate them to reduce the dimension of matrices and to save computational cost.
In the demonstrations shown in this paper, we truncate singular values smaller than $10^{-12}$.
To avoid overfitting the noise of the Green's function, we define the following $L_1$-norm regularized cost function
\begin{equation}
    L(\tiRvec) = \frac{1}{2}\sum_\ell \left(\tiG_\ell - S_\ell \tiR_\ell\right)^2 + \lambda \sum_\ell |\tiR_\ell|,
    \label{eq:L1}
\end{equation}
and transform the problem into the optimization problem as
\begin{equation}
    \tiRvec^* = \minarg_{\tiRvec} L(\tiRvec).
\end{equation}
The $L_1$-norm term (the second term) in $L$ removes very small (noisy) components and makes AC stable against the noise, as seen later.
With fixed $\lambda$, this optimization problem can be iteratively solved by using the alternating direction method of multipliers (ADMM)~\cite{BoydPCPE2011}.
Finding the optimal value of $\lambda$ is a problem remaining to be solved.
One way to solve it is elbow analysis of the error $\chi^2 = \sum_\ell (\tiG_\ell - S_\ell \tiR_\ell)^2/2$ [the first term in Eq.~(\ref{eq:L1})].
In this analysis, we solve the optimization problem at fixed $\lambda$ and calculate the error term $\chi^2(\lambda)$, and then plot $\chi^2(\lambda)$ against $\lambda$ in a log-log scale.
We can estimate the optimal value of $\lambda$ as the kink of the curve.
Once $\tiRvec^*$ is obtained, the solution of Eq.~(\ref{eq:disceq}), $\bm{\rho}^*$, is calculated as
\begin{equation}
    \rho_j^* = \sum_{\ell} V_{j\ell}\tiR^*_\ell.
\end{equation}

Since ADMM can be easily extended to deal with cost functions with more constraints,
we finally consider the following cost function,
\begin{equation}
\begin{split}
    L_\text{SpM}(\tiRvec) &= \frac{1}{2}\sum_\ell \left(\tiG_\ell - S_\ell \tiR_\ell\right)^2 + \lambda \sum_\ell |\tiR_\ell| \\
    &+ \lim_{v \to \infty} v \left(\rho_\Sigma - \sum_i \rho_i \right)^2
    + \lim_{\gamma \to \infty} \gamma \sum_i \Theta(-\rho_i),
\end{split}
\label{eq:cost-SpM}
\end{equation}
where the third and the fourth terms denote the sum-rule and the non-negativity of the spectrum function, respectively, and $\Theta(x)$ is the Heaviside step function.
The reader is referred to the appendix of Ref.~\cite{OtsukiOSY2017} for details of the ADMM algorithm for this cost function.

The SpM method also has its own advantages and disadvantages, as we show later.
This method gives us a very robust spectrum without any \textit{a priori} knowledge, such as the so-called default model in the maximum entropy method.
On the other hand, an unphysical oscillation appears in the obtained spectrum, especially in the low-frequency region.
This comes from the fact that the IR basis, which is oscillating in the frequency domain, will be truncated for noise reduction~\footnote{It should be noted that other methods adopting an oscillating basis such as Chebyshev polynomials may suffer from this problem.}.

\subsection{SpM-\Pade{} \ --- proposed method}
The \Pade method seems to give us a good estimation in regions where the estimation is robust against the noise.
In general, such a region lies near the origin of frequency, where the SpM method suffers from oscillation.
To overcome this problem,
we propose the SpM-\Pade{} method, for robust and smooth analytic continuation based on the \Pade and the SpM methods.
This method estimates the spectrum in a similar way to the SpM method, but puts an additional term into the cost function, Eq.~(\ref{eq:cost-SpM}), the distance from the spectrum estimated by the \Pade{} method.
Instead of explicitly specifying the region where the \Pade estimation is used in the final result, the SpM-\Pade method uses the precision of the \Pade{} estimation at each frequency as a weight, as seen later.

In this method, we first estimate the expectation values $\rho_i^\text{\Pade}$ and the variance $\left(\sigma_i^\text{\Pade}\right)^2$ by the \Pade approximation from independent $N_\text{pade}$ Green's functions generated by adding Gaussian noise into the original Green's function~\footnote{The \Pade method is very fast, and so the time to estimate $\rho^\text{\Pade}$ and $\sigma_i^\text{\Pade}$ is negligible.}.
Once these \Pade{} results are calculated, the cost function of the SpM-\Pade{} method is defined as the following:
\begin{equation}
    L_\text{SpM-\Pade}(\tiRvec) = L_\text{SpM}(\tiRvec)
    + \frac{\eta}{2}\sum_i w_i \left(\rho_i^\text{\Pade} - \rho_i \right)^2,
    \label{eq:cost-spmpade}
\end{equation}
where $L_\text{SpM}$ is the cost function in the SpM method, Eq.~(\ref{eq:cost-SpM}),
and $w_i$ is a weight determining how much the spectrum of the \Pade spectrum $\rho^\text{\Pade}$ is included in the final result.
We adopt the following simple form for the weight function:
\begin{equation}
w_i = \left[1 + \left(\frac{\sigma_i^\text{\Pade}}{\rho_i^\text{\Pade}}\right)^2\right]^{-1}.
\label{eq:weight}
\end{equation}

As for the original SpM method, once hyperparameters $\lambda$ and $\eta$ are given, the minimization problem with the cost function $L_\text{SpM-\Pade}$ can be solved with ADMM,
and then we should find the optimal values of $\lambda$ and $\eta$.
In this study, we fixed $\eta$ to 1 and decided the optimal value of $\lambda$ by elbow analysis as in the original SpM method.
The $\eta$ dependence of the resulting spectrum will be discussed later.

\section{Numerical results}
\label{sec:demonstration}
To demonstrate the SpM-\Pade method, we performed three benchmark tests:
(a) a test for a fermionic system with a symmetric spectrum, where the SpM method works well but the \Pade method does not (the same spectrum is tested in Refs.~\cite{OtsukiOSY2017, Yoshimi:2019bt}),
(b) a test for a fermionic system with a double-peak spectrum, where neither the \Pade nor the SpM method works well,
and (c) a test for the Hubbard model on the square lattice as a real-world example.

The procedure of benchmark tests (a) and (b) is as follows.
First, from the ``exact'' spectrum, we obtained the ``exact'' fermionic imaginary-time Green's function $G_\text{exact}(\tau)$ via Eq~(\ref{eq:inteq}) with $\beta = 100$, and generated 30 independent imaginary-time Green's functions as samples by adding Gaussian noise with deviation $\sigma$ to $G_\text{exact}$ at each imaginary time slice independently and individually.
The range of frequency is from $\omega_\text{min}=-4$ to $\omega_\text{max}=4$ and the numbers for the discretization of frequency and imaginary time are $N_\omega = 1001$ and $N_\tau = 4001$, respectively.
We next estimated the spectrum function from each sample, and then calculated the mean and the variance of 30 estimates at each frequency.
For test (c), we reconstructed the spectral function from the Matsubara Green's function calculated by the dynamical mean-field theory (DMFT)~\cite{RevModPhys.68.13} with the PIMC method.
We used an open-source software package \textit{SpM}~\cite{OtsukiOSY2017, Yoshimi:2019bt, SpM} for performing the \Pade method and the SpM method.
We also implemented the SpM-\Pade{} method based on \textit{SpM}.

\subsection{Three-peak spectrum}
\begin{figure*}
    \centering
    \includegraphics[width=\linewidth]{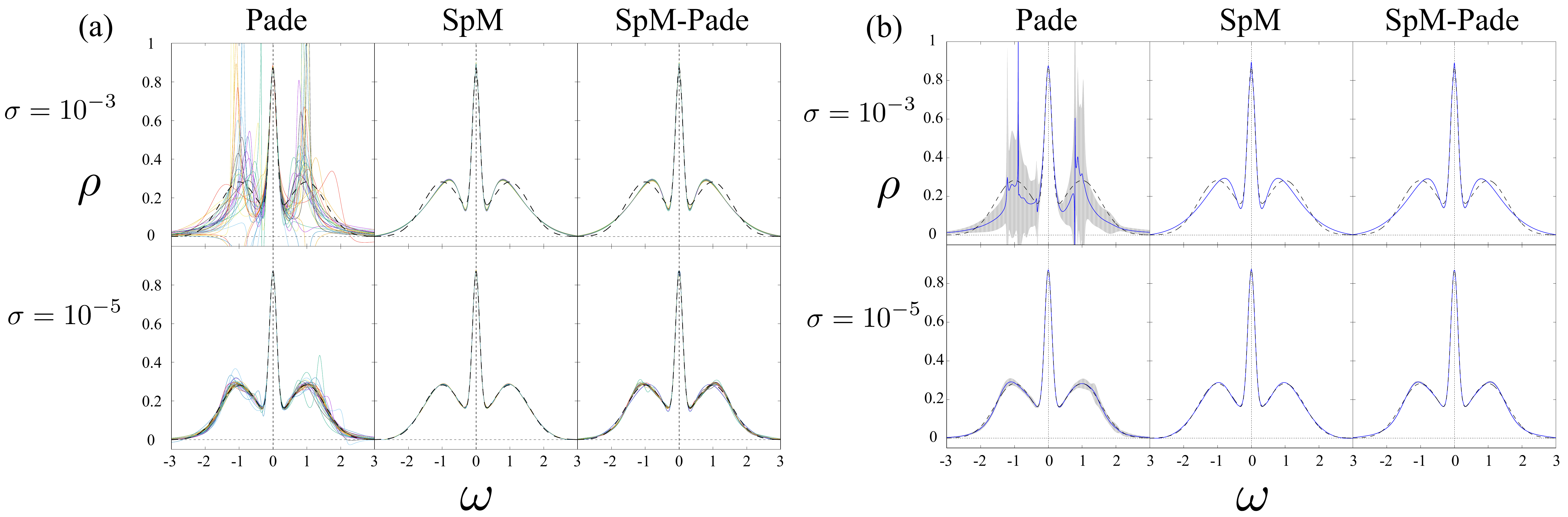}
    \caption{(a) Three-peak spectrum from 30 samples reconstructed by the \Pade (left), SpM (middle), and SpM-\Pade{} methods.
    The panels in the top row and the bottom row denote spectra from the Green's functions with large noise ($10^{-3}$) and small noise ($10^{-5}$), respectively.
    The exact spectrum is shown as the black dashed line.
    (b) Mean value (blue curve) and standard deviation (shaded area) of 30 reconstructions denoted in (a).
    }
    \label{fig:demo_three_peaks}
\end{figure*}

In this subsection, the ``exact'' spectrum is $\rho_\text{exact}(\omega) = 0.2p(\omega; \omega_0=0, \sigma^2=0.075) + 0.4p(\omega; \omega_0=1.0, \sigma^2=0.4) + 0.4p(\omega; \omega_0=-1.0, \sigma^2=0.4)$, where $p(\omega; \omega_0, \sigma^2)$ is a normalized Gauss distribution with the width $\sigma$ at the position $\omega_0$,
\begin{equation}
    p(\omega; \omega_0, \sigma^2) = \frac{1}{\sqrt{2\pi \sigma^2 }}e^{-\frac{(\omega-\omega_0)^2}{2\sigma^2}}.
\end{equation}
Figure~\ref{fig:demo_three_peaks} (a) shows
reconstructed spectra obtained by the three methods, \Pade{} (left column), SpM (middle column), and SpM-\Pade{} (right column), from 30 independent samples with two different noise levels, $\sigma = 10^{-3}$ (top row) and $\sigma = 10^{-5}$ (bottom row),
and (b) shows the means (line) and standard deviations (shaded region) of the 30 spectra.
Note that the minimum value of the absolute value of the exact Green's function is about $2.6\times10^{-2}$.
The black dashed curve denotes the ``exact'' spectrum $\rho_\text{exact}$.
In the case that the noise is small enough ($\sigma=10^{-5}$), all three methods give accurate and robust results as seen in the bottom panels.
In the large-noise case ($\sigma=10^{-3}$), on the other hand, the \Pade{} result becomes unstable in the second peak at $\omega = \pm 1$.
Although the \Pade{} method fails, the SpM method still obtains robust results.
It can be seen that SpM-\Pade{} also performed well, since the weight $w_i$ is automatically suppressed when the variance of the \Pade{} method becomes large as shown in Eq.(\ref{eq:weight}).

\subsection{Two-peak spectrum}
\begin{figure*}
    \centering
    \includegraphics[width=\linewidth]{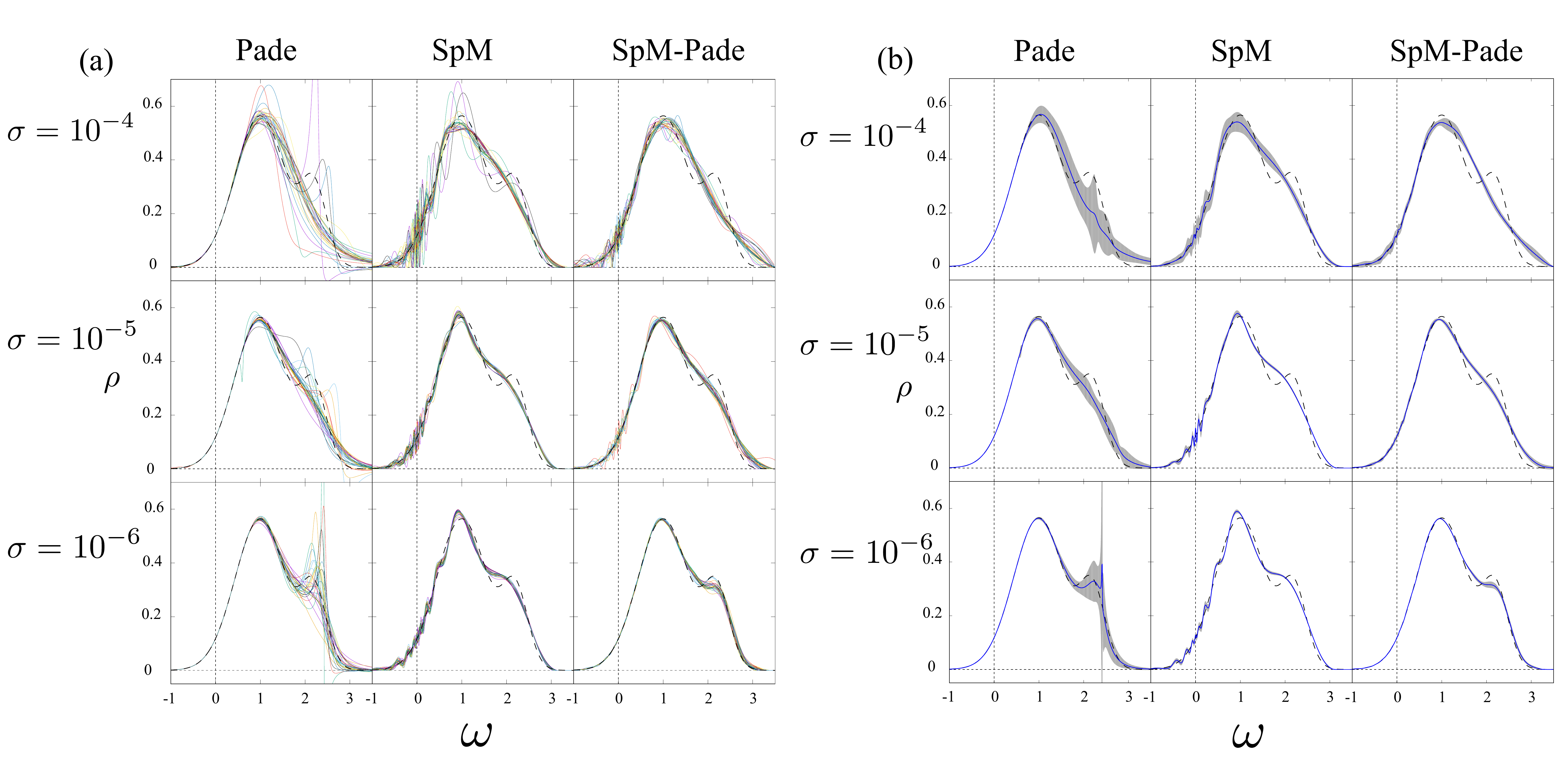}
    \caption{(a) Two-peak spectrum from 30 samples reconstructed by the \Pade (left), SpM (center), and SpM-\Pade (right) methods.
    The panels on the top, middle, and bottom row shows spectra from the Green's functions with large noise ($10^{-4})$, medium noise ($10^{-5}$), and small noise ($10^{-6}$), respectively.
    The exact spectrum is shown as the black dashed curve.
    (b) Mean value (blue curve) and standard deviation (shaded area) of 30 reconstructions.
    }
    \label{fig:demo_two_peaks}
\end{figure*}

In this subsection, the ``exact'' spectrum is $\rho_\text{exact}(\omega) = 0.8p(\omega; \omega_0=1, \sigma^2=0.4) + 0.2p(\omega; \omega_0=2.2, \sigma^2=0.2).$
Figure~\ref{fig:demo_two_peaks} (a) shows the spectra reconstructed by the three methods, \Pade{} (left column), SpM (middle column), and SpM-\Pade{} (right column), from 30 independent samples with three different noise levels, $\sigma = 10^{-4}$ (top row), $\sigma = 10^{-5}$ (middle row), and $\sigma = 10^{-6}$ (bottom row),
and panel (b) shows the means (line) and standard deviations (shaded region) of the 30 spectra.
Note that the minimum value of the absolute value of the exact Green's function is about $3.7\times10^{-3}$.

First, we will consider the \Pade result shown in the left panels.
Below the first peak at $\omega \lesssim 1$, the \Pade method gives a precise and robust estimation.
Around the second peak at $\omega \sim 2.2$, however, this becomes unstable against the noise of the input.
This is because this method consumes most of the information in $G(\tau)$ to restore the first peak.
The SpM result depicted in the middle column, on the other hand, is more robust as the narrower shaded region indicates.
At the low-frequency region $\omega \simeq 0$, however, the SpM result suffers from an artificial oscillation.
It is clearly shown in the right panels that the oscillation in the result of the SpM method vanishes in that of the SpM-\Pade method, and the robustness of the SpM method remains.

\begin{figure}
    \centering
    \includegraphics[width=\linewidth]{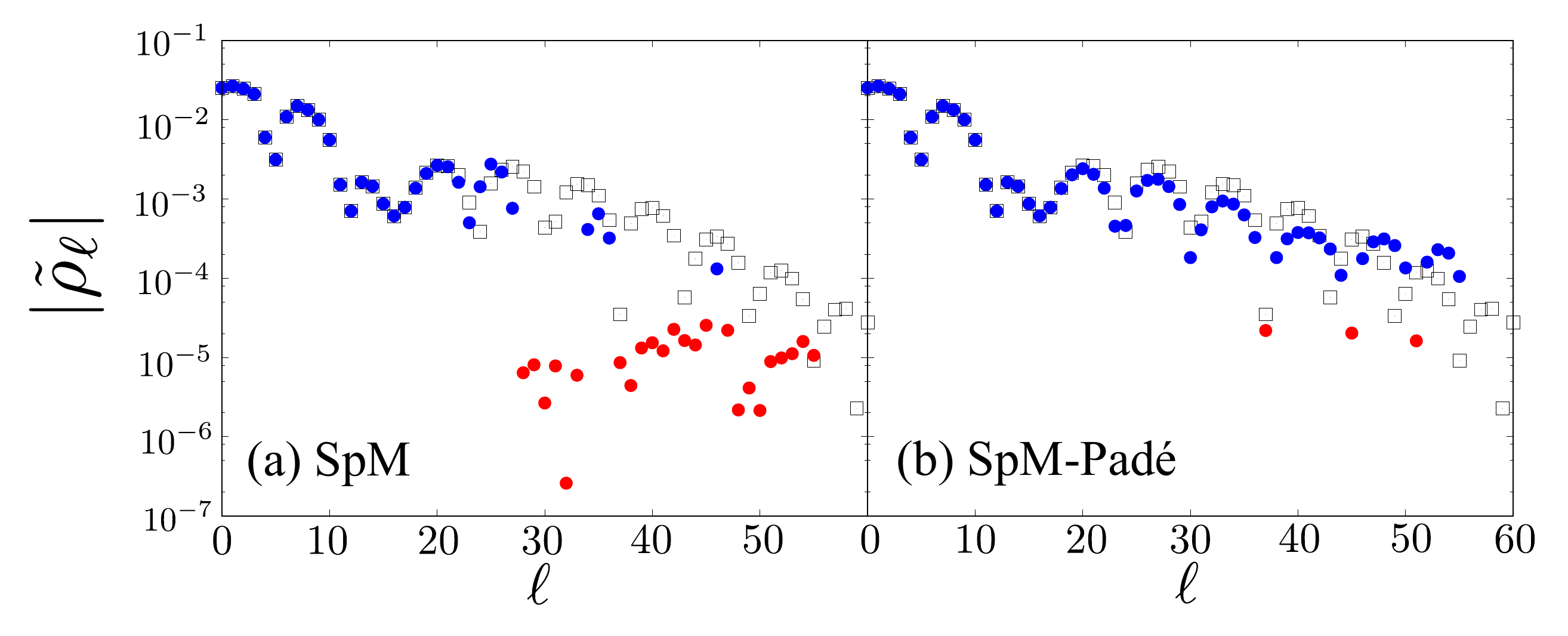}
    \caption{Spectrum function in the IR basis.
    Open squares denote components of the ``exact'' spectrum (see body text).
    Filled circles denote those of the reconstructed spectrum from one sample with $\sigma=10^{-6}$ by (a) the SpM method and (b) the SpM-\Pade{} method.
    The red symbols are removed by the ADMM algorithm and the blue ones remains.
    }
    \label{fig:rho_sv}
\end{figure}

To see the reason for the robustness and the oscillation in the SpM method, we first show the IR components of the spectra $\tilde{\rho}_{\ell}$ where $\ell$ is the index of the components in descending order in Fig. \ref{fig:rho_sv} (a).
The open squares are $\tilde{\rho}$ for the exact spectrum and
blue circles are $\tilde{\rho}$ for the reconstructed spectrum from one sample with $\sigma=10^{-6}$ noise.
The red circles stand for the components which are too small and are removed through the ADMM algorithm as noise.
We truncated the components with small singular values, $S_\ell < 10^{-12}$ ($\ell > 55$).
It is seen that $\tilde{\rho}_{\ell}$ tends to become exponentially smaller with increasing $\ell$ and the deviation from the exact spectrum becomes larger from the components having the same magnitude as the noise level.
The removal of these noisy components makes the SpM method robust against noise, 
but this also introduces some oscillation into the result as a truncation error.

\begin{figure}
    \centering
    \includegraphics[width=\linewidth]{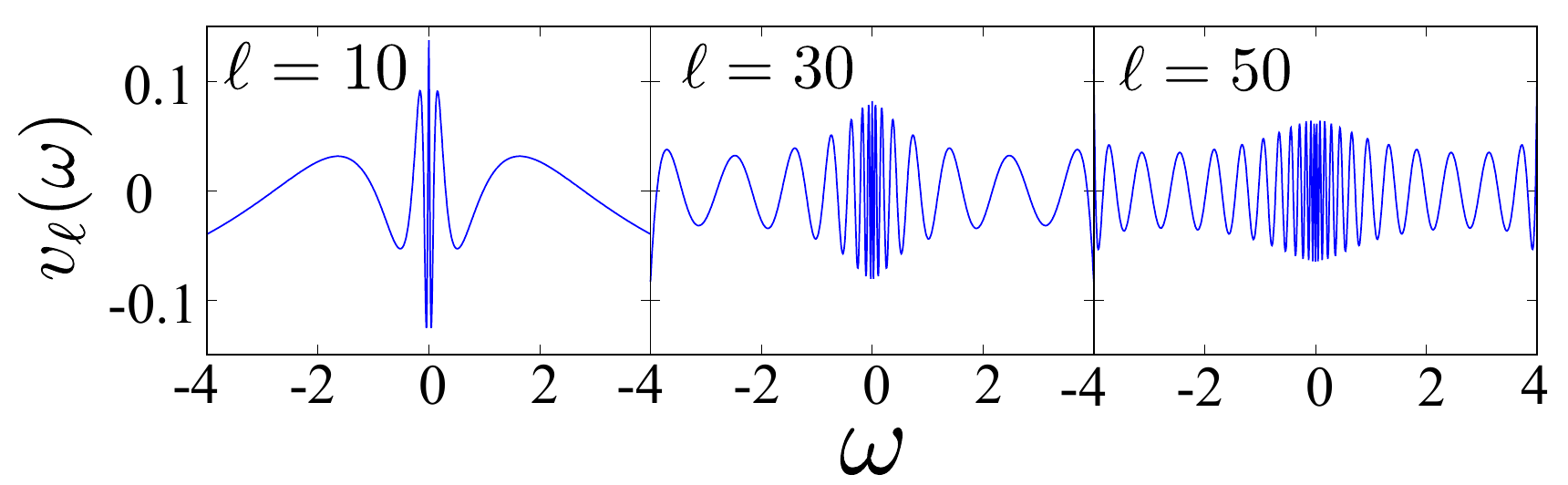}
    \caption{IR basis of spectrum $v_\ell(\omega_j) = V_{j\ell}$ in frequency space with fixed $\ell = 10$ (left), $30$ (middle), and $50$ (right) for $\beta = 100$.}
    \label{fig:VL}
\end{figure}

Figure \ref{fig:VL} shows some of the IR basis of the spectrum in frequency space, $v_\ell(\omega_j) = V_{j\ell}$ with fixed $\ell = 10, 30, $ and $50$.
Since these functions strongly oscillate around the origin $\omega \sim 0$, the spectrum function $\rho(\omega)$ obtained by the SpM method also oscillates in the low-frequency region as a truncation error.
Since the AC using the \Pade{} approximation is highly accurate in the low-frequency range, it is expected to protect information even in the high component part of $\ell$, which is strongly affected by noise.
Figure~\ref{fig:rho_sv} (b) shows the AC result using the SpM-\Pade{} method. It is seen that some of the removed IR components are restored, i.e., that this method succeeds in extracting correct information even from the components affected by noise.
This is why the SpM-\Pade{} method succeeds in removing the oscillation.

\subsection{Weight for \Pade}
\begin{figure}
    \centering
    \includegraphics[width=\linewidth]{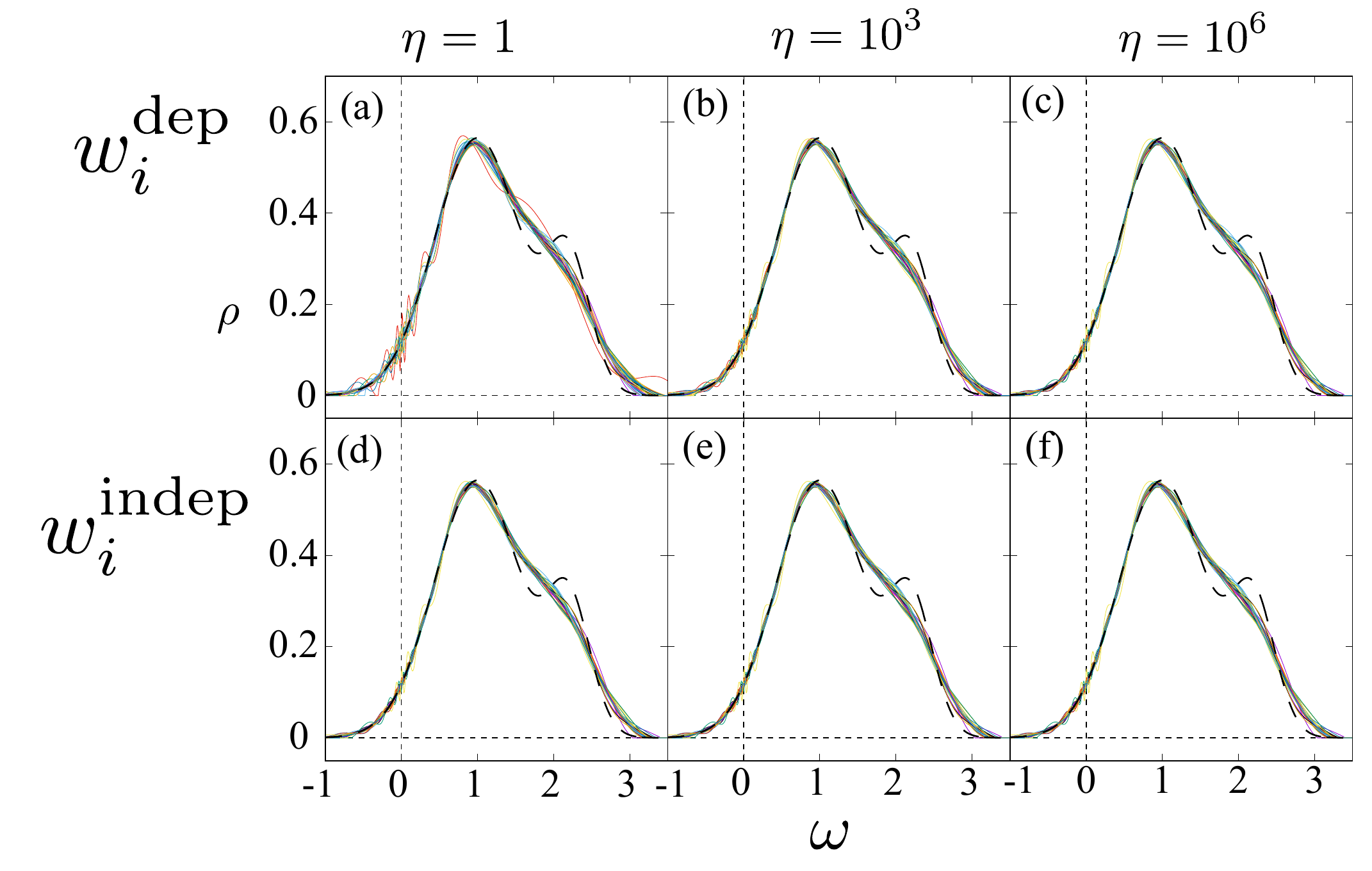}
    \caption{Two-peaks spectra from 30 samples with $\sigma=10^{-5}$ noise reconstructed by the SpM-\Pade method.
    The upper panels (a)--(c) depict the results obtained by adopting $w_i^\text{dep} = \left[1+\left(\sigma_i^\text{\Pade}/\rho_i^\text{\Pade}\right)^2\right]^{-1}$ for the \Pade weight function, while the lower ones (d)--(f) adopt $w_i^\text{indep} = 1$.
    The left, middle, and right panels show the result obtained by using $\eta=1, 10^3,$ and $10^6$, respectively.
    }
    \label{fig:eta_deps}
\end{figure}

In the demonstrations, we let the \Pade coefficient $\eta$ be 1 and adopted the frequency-dependent \Pade weight function $w_i$ (Eq.~(\ref{eq:weight})).
Finally, we examine the effect of the hyperparameter $\eta$ and the weight function $w_i$ by using the Green's function with noise of $\sigma = 10^{-5}$ from the two-peak spectrum (the same Green's functions are used in the middle row of Fig.~\ref{fig:demo_two_peaks} (a)).
Figure \ref{fig:eta_deps} shows the 30 two-peak spectra reconstructed by using the SpM-\Pade method with $\eta = 1$ (left), $10^3$ (middle), and $10^6$ (right).
The upper panels depict the results with frequency-dependent weight,
\begin{equation}
w_i = w_i^\text{dep} = \left[1 + \left(\frac{\sigma_i^\text{\Pade}}{\rho_i^\text{\Pade}}\right)^2\right]^{-1},
\label{eq:weight_dep}
\end{equation}
and the lower panels show the results with frequency-independent weight,
\begin{equation}
    w_i = w_i^\text{indep} = 1.
\end{equation}
The figure shows us the following:
(i) Panels (a) and (b) show that the increase of $\eta$ suppresses the oscillation in the spectrum near $\omega = 0$.
(ii) Wider regions of the frequency where the \Pade spectrum are included increases $\eta$ virtually (shown in panels (a), (b), and (d)).
(iii) $\eta=10^3$ and $\eta=10^6$ seem to result in the same spectrum.
This is because, while the large $\eta$ term favors the \Pade spectrum, the error in the Green's function, $\left(\tilde{G}_\ell - S_\ell \tilde{\rho}_\ell\right)^2$, disfavors the \Pade spectrum.

The cost functions $L_\text{SpM}$ and $L_\text{SpM-\Pade}$ are not suitable for the cost of optimizing hyperparameters, because they trivially take the minimum value, zero, when $\lambda = \eta = 0$ and $\tilde{\rho}_\ell = \tilde{G}_\ell / S_\ell$, and this results in overfitting.
This is one of the reasons why the elbow method is used in optimizing $\lambda$ under fixed $\eta$,
but the extension of this method to two or more hyperparameters is not straightforward.
The search for more sophisticated optimization methods is a future problem.

\subsection{Real example: Hubbard model on square lattice}
\label{sec:hubbard}
As an example of real calculations, we apply the SpM-\Pade{} method to PIMC data computed in the Hubbard model. The Hamiltonian is given by
\begin{equation}
\mathcal{H} = -t \sum_{\braket{ij}}\sum_{\sigma = \uparrow, \downarrow} \left[ \hat{c}_{i\sigma}^\dagger \hat{c}_{j\sigma} + \text{h.c.} \right]
+ U \sum_i \hat{n}_{i\uparrow}\hat{n}_{i\downarrow},
\end{equation}
where $\hat{c}_{i\sigma} (\hat{c}_{i\sigma}^{\dagger})$ is the annihilation (creation) operator of the electron with spin $\sigma$ on $i$th site, $\hat{n}_{i\sigma} \equiv \hat{c}_{i\sigma}^{\dagger}\hat{c}_{i\sigma}$ is the number operator, and $\sum_{\braket{ij}}$ denotes the summation over pairs of nearest neighbor sites.
We set the parameters at $U = 12$, $n=0.8$, and $\beta = 10$ in the unit of $t=1$ for the following reasons.
At half filling, $n=1$, the system becomes a Mott insulating state having a charge gap, because $U=12$ is large enough compared with the bandwidth $W=8$.
Therefore, the single-particle excitation spectrum $\rho(\omega)$ exhibits two peaks away from $\omega=0$, i.e., at $\omega=\pm U/2$.
When we dope the Mott state, an additional peak characterizing metallic states emerges around $\omega=0$, and thus a three-peak structure is expected in $\rho(\omega)$.
Such spectra realized due to strong correlations are difficult to reproduce by AC and are suitable for demonstration. Thus, we set $n=0.8$.

We computed the Matsubara Green's function by the DMFT combined with the PIMC method,
using open-source software packages.
Leaving its details to the Appendix,
here we only remark that the relative statistical errors $\left|\Delta G_i/G_i\right|$ of this calculation is about $0.05$, and hence it corresponds to the ``noisy'' cases in the other demonstrations.
In the AC procedure, the range of the frequency is $\omega \in [-15, 25]$, and the numbers of frequency points and imaginary time points are $N_\omega = 2001$ and $N_\tau = 10001$, respectively.
We used $w_i^\text{dep}$ in Eq.~(\ref{eq:weight_dep}) with $\eta = 10$ for the \Pade weight function.

\begin{figure}
    \centering
    \includegraphics[width=\linewidth]{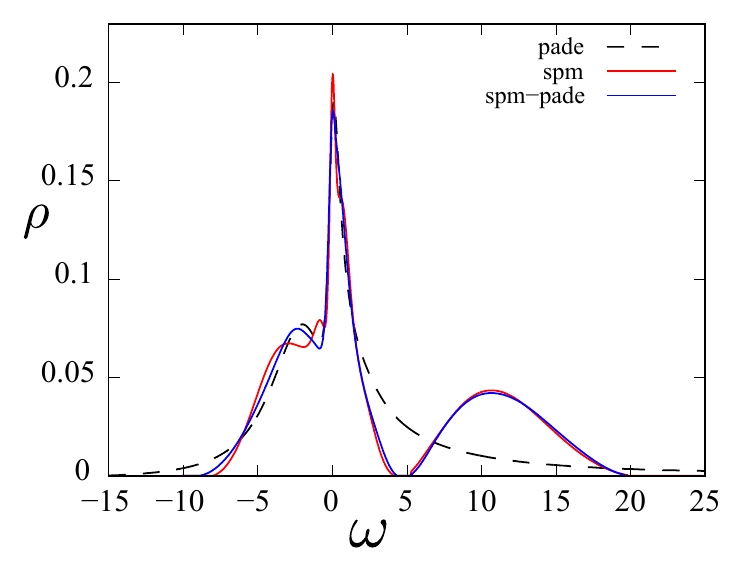}
    \caption{
    The single-particle excitation spectrum $\rho(\omega)$ in the Hubbard model calculated from the Green's function obtained by DMFT+PIMC under $U/t=12$, $n=0.8$, and $\beta=10$.
    The black dashed curve, the red curve, and the blue curve denote the spectra reconstructed by using the \Pade{} method, the SpM method, and the SpM-\Pade{} method, respectively.
    }
    \label{fig:demo_hubbard}
\end{figure}

Fig.~\ref{fig:demo_hubbard} shows $\rho(\omega)$ computed by AC using the \Pade (a black broken line), the SpM (a red line), and the SpM-\Pade{} (a blue line) methods.
The SpM result shows the upper Hubbard peak around $\omega=10$, while it is missing in the \Pade result. From the physical consideration as above, the upper Hubbard peak should be there and therefore the SpM spectrum is reasonable. The SpM-\Pade result inherits this feature.
Around $\omega=0$, on the other hand, the SpM method seems to suffer from an artificial oscillation, which is suppressed in the SpM-\Pade method.
The SpM-\Pade spectrum thus exhibits a physically reasonable spectrum in the whole frequency region.
This result demonstrates the advantage of our method, the SpM-\Pade, in real simulations.

\section{Summary}
\label{sec:summary}

In this paper, we focused on two methods for analytic continuation from the imaginary-time Green's function to the real-frequency spectral function.
The \Pade method and the SpM method show an example of the bias-variance trade-off: The \Pade method gives a low-bias but high-variance result, while the SpM method gives a low-variance but high-bias result.
The former is due to overfitting of the input, while the latter is due to the over-trimming of bases.
Combining \Pade with SpM, we recovered the bases trimmed in the SpM and acquired smooth and accurate low-frequency behavior, keeping the robustness of SpM.
As a result, our SpM-\Pade method achieves both low bias and low variance.

In addition, recently, computational techniques using the compact IR basis have been developed to calculate the dynamic susceptibility\cite{PhysRevB.97.205111,PhysRevB.102.134503,PhysRevResearch.3.033168}. It is expected that by using the basis obtained by SpM-\Pade{}, the accuracy of these analytic calculations will be improved. Applications of the method to these applied calculations are interesting challenges but are left as future issues.

\begin{acknowledgments}
We thank H. Shinaoka and Y. Nakanishi-Ohno for fruitful discussions.
YM and KY were supported by Building of Consortia for the Development of Human Resources in Science and Technology, MEXT, Japan. This work was supported by JSPS KAKENHI grants No. 19K03649, No. 20K20522, No. 21H01003, and No. 21H01041.
Some of the computation in this work has been done using the facilities of the Supercomputer Center, the Institute for Solid State Physics, the University of Tokyo.
\end{acknowledgments}

\appendix
% \section{Numerical conditions of DMFT+PIMC method used in Sec.~\ref{sec:hubbard}}
\section{Details of DMFT+PIMC calculations}
\label{sec:appendix}

In this appendix, we describe the details of the DMFT+PIMC calculation in Sec.~\ref{sec:hubbard}.
For the DMFT calculation, we used an open-source software package DCore version 3.0.0~\cite{dcore, dcore_doc} implemented on TRIQS library~\cite{triqs}.
The number of iterations is 40, and the mixing parameter for the self-energy is 0.5.
The time reversal symmetry is assumed and hence the average over the spin is taken.
For the PIMC calculation, we used an open-source software package ALPS/CTHYB-segment (with commit hash \texttt{623aa1a868})~\cite{Werner2006, Hafermann2013, ALPS-CTHYB-SEGMENT, ALPSCore2017, ALPSCore2018}.
The number of MC updates between measurements is 50, and the number of thermalization steps is $10^6$.
The number of measurements in the last iteration of the DMFT scheme was $1945813$\footnote{ALPS/CTHYB-segment has a parameter for specifying the maximum runtime of code, which is set as 60 second in this demonstration.}.
The ALPS/CTHYB-segment program is executed with 16 MPI processes + 8 OpenMP threads.

% \bibliography{main}% Produces the bibliography via BibTeX.

\begin{thebibliography}{45}%
\makeatletter
\providecommand \@ifxundefined [1]{%
 \@ifx{#1\undefined}
}%
\providecommand \@ifnum [1]{%
 \ifnum #1\expandafter \@firstoftwo
 \else \expandafter \@secondoftwo
 \fi
}%
\providecommand \@ifx [1]{%
 \ifx #1\expandafter \@firstoftwo
 \else \expandafter \@secondoftwo
 \fi
}%
\providecommand \natexlab [1]{#1}%
\providecommand \enquote  [1]{``#1''}%
\providecommand \bibnamefont  [1]{#1}%
\providecommand \bibfnamefont [1]{#1}%
\providecommand \citenamefont [1]{#1}%
\providecommand \href@noop [0]{\@secondoftwo}%
\providecommand \href [0]{\begingroup \@sanitize@url \@href}%
\providecommand \@href[1]{\@@startlink{#1}\@@href}%
\providecommand \@@href[1]{\endgroup#1\@@endlink}%
\providecommand \@sanitize@url [0]{\catcode `\\12\catcode `\$12\catcode
  `\&12\catcode `\#12\catcode `\^12\catcode `\_12\catcode `\%12\relax}%
\providecommand \@@startlink[1]{}%
\providecommand \@@endlink[0]{}%
\providecommand \url  [0]{\begingroup\@sanitize@url \@url }%
\providecommand \@url [1]{\endgroup\@href {#1}{\urlprefix }}%
\providecommand \urlprefix  [0]{URL }%
\providecommand \Eprint [0]{\href }%
\providecommand \doibase [0]{https://doi.org/}%
\providecommand \selectlanguage [0]{\@gobble}%
\providecommand \bibinfo  [0]{\@secondoftwo}%
\providecommand \bibfield  [0]{\@secondoftwo}%
\providecommand \translation [1]{[#1]}%
\providecommand \BibitemOpen [0]{}%
\providecommand \bibitemStop [0]{}%
\providecommand \bibitemNoStop [0]{.\EOS\space}%
\providecommand \EOS [0]{\spacefactor3000\relax}%
\providecommand \BibitemShut  [1]{\csname bibitem#1\endcsname}%
\let\auto@bib@innerbib\@empty
%</preamble>
\bibitem [{\citenamefont {Abrikosov}\ \emph {et~al.}(1963)\citenamefont
  {Abrikosov}, \citenamefont {Gorkov},\ and\ \citenamefont
  {Dzyaloshinski}}]{AbrikosovGD1963}%
  \BibitemOpen
  \bibfield  {author} {\bibinfo {author} {\bibfnamefont {A.}~\bibnamefont
  {Abrikosov}}, \bibinfo {author} {\bibfnamefont {L.}~\bibnamefont {Gorkov}},\
  and\ \bibinfo {author} {\bibfnamefont {I.}~\bibnamefont {Dzyaloshinski}},\
  }\href@noop {} {\emph {\bibinfo {title} {{Methods of quantum field theory in
  statistical physics}}}}\ (\bibinfo  {publisher} {Dover},\ \bibinfo {address}
  {New York, N.Y.},\ \bibinfo {year} {1963})\BibitemShut {NoStop}%
\bibitem [{\citenamefont {Gull}\ \emph {et~al.}(2011)\citenamefont {Gull},
  \citenamefont {Millis}, \citenamefont {Lichtenstein}, \citenamefont
  {Rubtsov}, \citenamefont {Troyer},\ and\ \citenamefont
  {Werner}}]{GullMLRTW2011}%
  \BibitemOpen
  \bibfield  {author} {\bibinfo {author} {\bibfnamefont {E.}~\bibnamefont
  {Gull}}, \bibinfo {author} {\bibfnamefont {A.~J.}\ \bibnamefont {Millis}},
  \bibinfo {author} {\bibfnamefont {A.~I.}\ \bibnamefont {Lichtenstein}},
  \bibinfo {author} {\bibfnamefont {A.~N.}\ \bibnamefont {Rubtsov}}, \bibinfo
  {author} {\bibfnamefont {M.}~\bibnamefont {Troyer}},\ and\ \bibinfo {author}
  {\bibfnamefont {P.}~\bibnamefont {Werner}},\ }\href
  {https://doi.org/10.1103/RevModPhys.83.349} {\bibfield  {journal} {\bibinfo
  {journal} {Rev. Mod. Phys.}\ }\textbf {\bibinfo {volume} {83}},\ \bibinfo
  {pages} {349} (\bibinfo {year} {2011})}\BibitemShut {NoStop}%
\bibitem [{\citenamefont {Gubernatis}\ \emph {et~al.}(2016)\citenamefont
  {Gubernatis}, \citenamefont {Kawashima},\ and\ \citenamefont
  {Werner}}]{GubernatisKW2016}%
  \BibitemOpen
  \bibfield  {author} {\bibinfo {author} {\bibfnamefont {J.}~\bibnamefont
  {Gubernatis}}, \bibinfo {author} {\bibfnamefont {N.}~\bibnamefont
  {Kawashima}},\ and\ \bibinfo {author} {\bibfnamefont {P.}~\bibnamefont
  {Werner}},\ }\href {https://doi.org/10.1017/CBO9780511902581} {\emph
  {\bibinfo {title} {Quantum Monte Carlo Methods: Algorithms for Lattice
  Models}}}\ (\bibinfo  {publisher} {Cambridge University Press},\ \bibinfo
  {year} {2016})\BibitemShut {NoStop}%
\bibitem [{\citenamefont {Schwandt}\ \emph {et~al.}(2009)\citenamefont
  {Schwandt}, \citenamefont {Alet},\ and\ \citenamefont
  {Capponi}}]{SchwandtAC2009}%
  \BibitemOpen
  \bibfield  {author} {\bibinfo {author} {\bibfnamefont {D.}~\bibnamefont
  {Schwandt}}, \bibinfo {author} {\bibfnamefont {F.}~\bibnamefont {Alet}},\
  and\ \bibinfo {author} {\bibfnamefont {S.}~\bibnamefont {Capponi}},\ }\href
  {https://doi.org/10.1103/PhysRevLett.103.170501} {\bibfield  {journal}
  {\bibinfo  {journal} {Phys. Rev. Lett.}\ }\textbf {\bibinfo {volume} {103}},\
  \bibinfo {pages} {170501} (\bibinfo {year} {2009})}\BibitemShut {NoStop}%
\bibitem [{\citenamefont {Wang}\ \emph {et~al.}(2015)\citenamefont {Wang},
  \citenamefont {Liu}, \citenamefont {Imri\ifmmode~\check{s}\else
  \v{s}\fi{}ka}, \citenamefont {Ma},\ and\ \citenamefont
  {Troyer}}]{WangLIMT2015}%
  \BibitemOpen
  \bibfield  {author} {\bibinfo {author} {\bibfnamefont {L.}~\bibnamefont
  {Wang}}, \bibinfo {author} {\bibfnamefont {Y.-H.}\ \bibnamefont {Liu}},
  \bibinfo {author} {\bibfnamefont {J.}~\bibnamefont
  {Imri\ifmmode~\check{s}\else \v{s}\fi{}ka}}, \bibinfo {author} {\bibfnamefont
  {P.~N.}\ \bibnamefont {Ma}},\ and\ \bibinfo {author} {\bibfnamefont
  {M.}~\bibnamefont {Troyer}},\ }\href
  {https://doi.org/10.1103/PhysRevX.5.031007} {\bibfield  {journal} {\bibinfo
  {journal} {Phys. Rev. X}\ }\textbf {\bibinfo {volume} {5}},\ \bibinfo {pages}
  {031007} (\bibinfo {year} {2015})}\BibitemShut {NoStop}%
\bibitem [{\citenamefont {Kolodrubetz}(2014)}]{Kolodrubetz2014}%
  \BibitemOpen
  \bibfield  {author} {\bibinfo {author} {\bibfnamefont {M.}~\bibnamefont
  {Kolodrubetz}},\ }\href {https://doi.org/10.1103/PhysRevB.89.045107}
  {\bibfield  {journal} {\bibinfo  {journal} {Phys. Rev. B}\ }\textbf {\bibinfo
  {volume} {89}},\ \bibinfo {pages} {045107} (\bibinfo {year}
  {2014})}\BibitemShut {NoStop}%
\bibitem [{\citenamefont {Motoyama}\ and\ \citenamefont
  {Todo}(2013)}]{MotoyamaT2013}%
  \BibitemOpen
  \bibfield  {author} {\bibinfo {author} {\bibfnamefont {Y.}~\bibnamefont
  {Motoyama}}\ and\ \bibinfo {author} {\bibfnamefont {S.}~\bibnamefont
  {Todo}},\ }\href {https://doi.org/10.1103/PhysRevE.87.021301} {\bibfield
  {journal} {\bibinfo  {journal} {Phys. Rev. E}\ }\textbf {\bibinfo {volume}
  {87}},\ \bibinfo {pages} {021301(R)} (\bibinfo {year} {2013})}\BibitemShut
  {NoStop}%
\bibitem [{\citenamefont {Motoyama}\ and\ \citenamefont
  {Todo}(2018)}]{MotoyamaT2018}%
  \BibitemOpen
  \bibfield  {author} {\bibinfo {author} {\bibfnamefont {Y.}~\bibnamefont
  {Motoyama}}\ and\ \bibinfo {author} {\bibfnamefont {S.}~\bibnamefont
  {Todo}},\ }\href {https://doi.org/10.1103/PhysRevB.98.195127} {\bibfield
  {journal} {\bibinfo  {journal} {Phys. Rev. B}\ }\textbf {\bibinfo {volume}
  {98}},\ \bibinfo {pages} {195127} (\bibinfo {year} {2018})}\BibitemShut
  {NoStop}%
\bibitem [{\citenamefont {{H. J. Vidberg }}\ and\ \citenamefont
  {Serene}(1977)}]{VidbergS1977}%
  \BibitemOpen
  \bibfield  {author} {\bibinfo {author} {\bibnamefont {{H. J. Vidberg }}}\
  and\ \bibinfo {author} {\bibfnamefont {J.~W.}\ \bibnamefont {Serene}},\
  }\href@noop {} {\bibfield  {journal} {\bibinfo  {journal} {Journal of Low
  Temperature Physics}\ }\textbf {\bibinfo {volume} {29}},\ \bibinfo {pages}
  {179} (\bibinfo {year} {1977})}\BibitemShut {NoStop}%
\bibitem [{\citenamefont {Kiss}(2019)}]{Kiss2019}%
  \BibitemOpen
  \bibfield  {author} {\bibinfo {author} {\bibfnamefont {A.}~\bibnamefont
  {Kiss}},\ }\href {https://doi.org/10.1103/physrevb.100.214417} {\bibfield
  {journal} {\bibinfo  {journal} {Physical Review B}\ }\textbf {\bibinfo
  {volume} {100}},\ \bibinfo {pages} {214417} (\bibinfo {year}
  {2019})}\BibitemShut {NoStop}%
\bibitem [{\citenamefont {Weh}\ \emph {et~al.}(2020)\citenamefont {Weh},
  \citenamefont {Otsuki}, \citenamefont {Schnait}, \citenamefont {Evertz},
  \citenamefont {Eckern}, \citenamefont {Lichtenstein},\ and\ \citenamefont
  {Chioncel}}]{Weh2020}%
  \BibitemOpen
  \bibfield  {author} {\bibinfo {author} {\bibfnamefont {A.}~\bibnamefont
  {Weh}}, \bibinfo {author} {\bibfnamefont {J.}~\bibnamefont {Otsuki}},
  \bibinfo {author} {\bibfnamefont {H.}~\bibnamefont {Schnait}}, \bibinfo
  {author} {\bibfnamefont {H.~G.}\ \bibnamefont {Evertz}}, \bibinfo {author}
  {\bibfnamefont {U.}~\bibnamefont {Eckern}}, \bibinfo {author} {\bibfnamefont
  {A.~I.}\ \bibnamefont {Lichtenstein}},\ and\ \bibinfo {author} {\bibfnamefont
  {L.}~\bibnamefont {Chioncel}},\ }\href
  {https://doi.org/10.1103/PhysRevResearch.2.043263} {\bibfield  {journal}
  {\bibinfo  {journal} {Phys. Rev. Research}\ }\textbf {\bibinfo {volume}
  {2}},\ \bibinfo {pages} {043263} (\bibinfo {year} {2020})}\BibitemShut
  {NoStop}%
\bibitem [{\citenamefont {Silver}\ \emph {et~al.}(1990)\citenamefont {Silver},
  \citenamefont {Sivia},\ and\ \citenamefont {Gubernatis}}]{SilverSG1990}%
  \BibitemOpen
  \bibfield  {author} {\bibinfo {author} {\bibfnamefont {R.~N.}\ \bibnamefont
  {Silver}}, \bibinfo {author} {\bibfnamefont {D.~S.}\ \bibnamefont {Sivia}},\
  and\ \bibinfo {author} {\bibfnamefont {J.~E.}\ \bibnamefont {Gubernatis}},\
  }\href {https://doi.org/10.1103/PhysRevB.41.2380} {\bibfield  {journal}
  {\bibinfo  {journal} {Phys. Rev. B}\ }\textbf {\bibinfo {volume} {41}},\
  \bibinfo {pages} {2380} (\bibinfo {year} {1990})}\BibitemShut {NoStop}%
\bibitem [{\citenamefont {Jarrell}\ and\ \citenamefont
  {Gubernatis}(1996)}]{JarrelG1996}%
  \BibitemOpen
  \bibfield  {author} {\bibinfo {author} {\bibfnamefont {M.}~\bibnamefont
  {Jarrell}}\ and\ \bibinfo {author} {\bibfnamefont {J.}~\bibnamefont
  {Gubernatis}},\ }\href
  {https://doi.org/https://doi.org/10.1016/0370-1573(95)00074-7} {\bibfield
  {journal} {\bibinfo  {journal} {Physics Reports}\ }\textbf {\bibinfo {volume}
  {269}},\ \bibinfo {pages} {133 } (\bibinfo {year} {1996})}\BibitemShut
  {NoStop}%
\bibitem [{\citenamefont {Yoon}\ \emph {et~al.}(2018)\citenamefont {Yoon},
  \citenamefont {Sim},\ and\ \citenamefont {Han}}]{YoonSH2018}%
  \BibitemOpen
  \bibfield  {author} {\bibinfo {author} {\bibfnamefont {H.}~\bibnamefont
  {Yoon}}, \bibinfo {author} {\bibfnamefont {J.-H.}\ \bibnamefont {Sim}},\ and\
  \bibinfo {author} {\bibfnamefont {M.~J.}\ \bibnamefont {Han}},\ }\href
  {https://doi.org/10.1103/PhysRevB.98.245101} {\bibfield  {journal} {\bibinfo
  {journal} {Phys. Rev. B}\ }\textbf {\bibinfo {volume} {98}},\ \bibinfo
  {pages} {245101} (\bibinfo {year} {2018})}\BibitemShut {NoStop}%
\bibitem [{\citenamefont {Fournier}\ \emph {et~al.}(2020)\citenamefont
  {Fournier}, \citenamefont {Wang}, \citenamefont {Yazyev},\ and\ \citenamefont
  {Wu}}]{FournierWYW2020}%
  \BibitemOpen
  \bibfield  {author} {\bibinfo {author} {\bibfnamefont {R.}~\bibnamefont
  {Fournier}}, \bibinfo {author} {\bibfnamefont {L.}~\bibnamefont {Wang}},
  \bibinfo {author} {\bibfnamefont {O.~V.}\ \bibnamefont {Yazyev}},\ and\
  \bibinfo {author} {\bibfnamefont {Q.~S.}\ \bibnamefont {Wu}},\ }\href
  {https://doi.org/10.1103/physrevlett.124.056401} {\bibfield  {journal}
  {\bibinfo  {journal} {Physical Review Letters}\ }\textbf {\bibinfo {volume}
  {124}},\ \bibinfo {pages} {056401} (\bibinfo {year} {2020})}\BibitemShut
  {NoStop}%
\bibitem [{\citenamefont {Kades}\ \emph {et~al.}(2020)\citenamefont {Kades},
  \citenamefont {Pawlowski}, \citenamefont {Rothkopf}, \citenamefont
  {Scherzer}, \citenamefont {Urban}, \citenamefont {Wetzel}, \citenamefont
  {Wink},\ and\ \citenamefont {Ziegler}}]{KadesPRSUWWZ2020}%
  \BibitemOpen
  \bibfield  {author} {\bibinfo {author} {\bibfnamefont {L.}~\bibnamefont
  {Kades}}, \bibinfo {author} {\bibfnamefont {J.~M.}\ \bibnamefont
  {Pawlowski}}, \bibinfo {author} {\bibfnamefont {A.}~\bibnamefont {Rothkopf}},
  \bibinfo {author} {\bibfnamefont {M.}~\bibnamefont {Scherzer}}, \bibinfo
  {author} {\bibfnamefont {J.~M.}\ \bibnamefont {Urban}}, \bibinfo {author}
  {\bibfnamefont {S.~J.}\ \bibnamefont {Wetzel}}, \bibinfo {author}
  {\bibfnamefont {N.}~\bibnamefont {Wink}},\ and\ \bibinfo {author}
  {\bibfnamefont {F.~P.~G.}\ \bibnamefont {Ziegler}},\ }\href
  {https://doi.org/10.1103/physrevd.102.096001} {\bibfield  {journal} {\bibinfo
   {journal} {Physical Review D}\ }\textbf {\bibinfo {volume} {102}},\ \bibinfo
  {pages} {096001} (\bibinfo {year} {2020})}\BibitemShut {NoStop}%
\bibitem [{\citenamefont {Xie}\ \emph {et~al.}(2021)\citenamefont {Xie},
  \citenamefont {Bao}, \citenamefont {Maier},\ and\ \citenamefont
  {Webster}}]{XieBMW2021}%
  \BibitemOpen
  \bibfield  {author} {\bibinfo {author} {\bibfnamefont {X.}~\bibnamefont
  {Xie}}, \bibinfo {author} {\bibfnamefont {F.}~\bibnamefont {Bao}}, \bibinfo
  {author} {\bibfnamefont {T.}~\bibnamefont {Maier}},\ and\ \bibinfo {author}
  {\bibfnamefont {C.}~\bibnamefont {Webster}},\ }\bibfield  {journal} {\bibinfo
   {journal} {Discrete {\&} Continuous Dynamical Systems - S}\ }\href
  {https://doi.org/10.3934/dcdss.2021088} {10.3934/dcdss.2021088} (\bibinfo
  {year} {2021})\BibitemShut {NoStop}%
\bibitem [{\citenamefont {Sandvik}(1998)}]{Sandvik1998}%
  \BibitemOpen
  \bibfield  {author} {\bibinfo {author} {\bibfnamefont {A.~W.}\ \bibnamefont
  {Sandvik}},\ }\href {https://doi.org/10.1103/PhysRevB.57.10287} {\bibfield
  {journal} {\bibinfo  {journal} {Phys. Rev. B}\ }\textbf {\bibinfo {volume}
  {57}},\ \bibinfo {pages} {10287} (\bibinfo {year} {1998})}\BibitemShut
  {NoStop}%
\bibitem [{\citenamefont {Mishchenko}\ \emph {et~al.}(2000)\citenamefont
  {Mishchenko}, \citenamefont {Prokof'ev}, \citenamefont {Sakamoto},\ and\
  \citenamefont {Svistunov}}]{MishchenkoPSS2000}%
  \BibitemOpen
  \bibfield  {author} {\bibinfo {author} {\bibfnamefont {A.~S.}\ \bibnamefont
  {Mishchenko}}, \bibinfo {author} {\bibfnamefont {N.~V.}\ \bibnamefont
  {Prokof'ev}}, \bibinfo {author} {\bibfnamefont {A.}~\bibnamefont
  {Sakamoto}},\ and\ \bibinfo {author} {\bibfnamefont {B.~V.}\ \bibnamefont
  {Svistunov}},\ }\href {https://doi.org/10.1103/PhysRevB.62.6317} {\bibfield
  {journal} {\bibinfo  {journal} {Phys. Rev. B}\ }\textbf {\bibinfo {volume}
  {62}},\ \bibinfo {pages} {6317} (\bibinfo {year} {2000})}\BibitemShut
  {NoStop}%
\bibitem [{\citenamefont {Beach}(2004)}]{Beach2004}%
  \BibitemOpen
  \bibfield  {author} {\bibinfo {author} {\bibfnamefont {K.~S.~D.}\
  \bibnamefont {Beach}},\ }\href@noop {} {\bibinfo {title} {Identifying the
  maximum entropy method as a special limit of stochastic analytic
  continuation}} (\bibinfo {year} {2004}),\ \Eprint
  {https://arxiv.org/abs/arXiv:cond-mat/0403055} {arXiv:cond-mat/0403055}
  \BibitemShut {NoStop}%
\bibitem [{\citenamefont {Fuchs}\ \emph {et~al.}(2010)\citenamefont {Fuchs},
  \citenamefont {Pruschke},\ and\ \citenamefont {Jarrell}}]{FuchsPJ2010}%
  \BibitemOpen
  \bibfield  {author} {\bibinfo {author} {\bibfnamefont {S.}~\bibnamefont
  {Fuchs}}, \bibinfo {author} {\bibfnamefont {T.}~\bibnamefont {Pruschke}},\
  and\ \bibinfo {author} {\bibfnamefont {M.}~\bibnamefont {Jarrell}},\ }\href
  {https://doi.org/10.1103/PhysRevE.81.056701} {\bibfield  {journal} {\bibinfo
  {journal} {Phys. Rev. E}\ }\textbf {\bibinfo {volume} {81}},\ \bibinfo
  {pages} {056701} (\bibinfo {year} {2010})}\BibitemShut {NoStop}%
\bibitem [{\citenamefont {Sandvik}(2016)}]{Sandvik2016}%
  \BibitemOpen
  \bibfield  {author} {\bibinfo {author} {\bibfnamefont {A.~W.}\ \bibnamefont
  {Sandvik}},\ }\href {https://doi.org/10.1103/PhysRevE.94.063308} {\bibfield
  {journal} {\bibinfo  {journal} {Phys. Rev. E}\ }\textbf {\bibinfo {volume}
  {94}},\ \bibinfo {pages} {063308} (\bibinfo {year} {2016})}\BibitemShut
  {NoStop}%
\bibitem [{\citenamefont {Shao}\ \emph {et~al.}(2017)\citenamefont {Shao},
  \citenamefont {Qin}, \citenamefont {Capponi}, \citenamefont {Chesi},
  \citenamefont {Meng},\ and\ \citenamefont {Sandvik}}]{ShaoQCCMS2017}%
  \BibitemOpen
  \bibfield  {author} {\bibinfo {author} {\bibfnamefont {H.}~\bibnamefont
  {Shao}}, \bibinfo {author} {\bibfnamefont {Y.~Q.}\ \bibnamefont {Qin}},
  \bibinfo {author} {\bibfnamefont {S.}~\bibnamefont {Capponi}}, \bibinfo
  {author} {\bibfnamefont {S.}~\bibnamefont {Chesi}}, \bibinfo {author}
  {\bibfnamefont {Z.~Y.}\ \bibnamefont {Meng}},\ and\ \bibinfo {author}
  {\bibfnamefont {A.~W.}\ \bibnamefont {Sandvik}},\ }\href@noop {} {\bibfield
  {journal} {\bibinfo  {journal} {Phys. Rev. X}\ }\textbf {\bibinfo {volume}
  {7}},\ \bibinfo {pages} {041072} (\bibinfo {year} {2017})}\BibitemShut
  {NoStop}%
\bibitem [{\citenamefont {Otsuki}\ \emph {et~al.}(2017)\citenamefont {Otsuki},
  \citenamefont {Ohzeki}, \citenamefont {Shinaoka},\ and\ \citenamefont
  {Yoshimi}}]{OtsukiOSY2017}%
  \BibitemOpen
  \bibfield  {author} {\bibinfo {author} {\bibfnamefont {J.}~\bibnamefont
  {Otsuki}}, \bibinfo {author} {\bibfnamefont {M.}~\bibnamefont {Ohzeki}},
  \bibinfo {author} {\bibfnamefont {H.}~\bibnamefont {Shinaoka}},\ and\
  \bibinfo {author} {\bibfnamefont {K.}~\bibnamefont {Yoshimi}},\ }\href
  {https://doi.org/10.1103/PhysRevE.95.061302} {\bibfield  {journal} {\bibinfo
  {journal} {Phys. Rev. E}\ }\textbf {\bibinfo {volume} {95}},\ \bibinfo
  {pages} {061302(R)} (\bibinfo {year} {2017})}\BibitemShut {NoStop}%
\bibitem [{\citenamefont {Otsuki}\ \emph {et~al.}(2020)\citenamefont {Otsuki},
  \citenamefont {Ohzeki}, \citenamefont {Shinaoka},\ and\ \citenamefont
  {Yoshimi}}]{SpMReview}%
  \BibitemOpen
  \bibfield  {author} {\bibinfo {author} {\bibfnamefont {J.}~\bibnamefont
  {Otsuki}}, \bibinfo {author} {\bibfnamefont {M.}~\bibnamefont {Ohzeki}},
  \bibinfo {author} {\bibfnamefont {H.}~\bibnamefont {Shinaoka}},\ and\
  \bibinfo {author} {\bibfnamefont {K.}~\bibnamefont {Yoshimi}},\ }\href
  {https://doi.org/10.7566/JPSJ.89.012001} {\bibfield  {journal} {\bibinfo
  {journal} {J.Phys. Soc. Jpn.}\ }\textbf {\bibinfo {volume} {89}},\ \bibinfo
  {pages} {012001} (\bibinfo {year} {2020})}\BibitemShut {NoStop}%
\bibitem [{Note1()}]{Note1}%
  \BibitemOpen
  \bibinfo {note} {Note that other methods using basis transformation and
  truncation also face similar oscillation. Oscillation in the SpM method,
  however, seems larger than that in others.}\BibitemShut {Stop}%
\bibitem [{\citenamefont {Shinaoka}\ \emph {et~al.}(2017)\citenamefont
  {Shinaoka}, \citenamefont {Otsuki}, \citenamefont {Ohzeki},\ and\
  \citenamefont {Yoshimi}}]{ShinaokaOOY2017}%
  \BibitemOpen
  \bibfield  {author} {\bibinfo {author} {\bibfnamefont {H.}~\bibnamefont
  {Shinaoka}}, \bibinfo {author} {\bibfnamefont {J.}~\bibnamefont {Otsuki}},
  \bibinfo {author} {\bibfnamefont {M.}~\bibnamefont {Ohzeki}},\ and\ \bibinfo
  {author} {\bibfnamefont {K.}~\bibnamefont {Yoshimi}},\ }\href
  {https://doi.org/10.1103/PhysRevB.96.035147} {\bibfield  {journal} {\bibinfo
  {journal} {Phys. Rev. B}\ }\textbf {\bibinfo {volume} {96}},\ \bibinfo
  {pages} {035147} (\bibinfo {year} {2017})}\BibitemShut {NoStop}%
\bibitem [{\citenamefont {Boyd}\ \emph {et~al.}(2011)\citenamefont {Boyd},
  \citenamefont {Parikh}, \citenamefont {Chu}, \citenamefont {Peleato},\ and\
  \citenamefont {Eckstein}}]{BoydPCPE2011}%
  \BibitemOpen
  \bibfield  {author} {\bibinfo {author} {\bibfnamefont {S.}~\bibnamefont
  {Boyd}}, \bibinfo {author} {\bibfnamefont {N.}~\bibnamefont {Parikh}},
  \bibinfo {author} {\bibfnamefont {E.}~\bibnamefont {Chu}}, \bibinfo {author}
  {\bibfnamefont {B.}~\bibnamefont {Peleato}},\ and\ \bibinfo {author}
  {\bibfnamefont {J.}~\bibnamefont {Eckstein}},\ }\href
  {https://doi.org/10.1561/2200000016} {\bibfield  {journal} {\bibinfo
  {journal} {Foundations and Trends® in Machine Learning}\ }\textbf {\bibinfo
  {volume} {3}},\ \bibinfo {pages} {1} (\bibinfo {year} {2011})}\BibitemShut
  {NoStop}%
\bibitem [{Note2()}]{Note2}%
  \BibitemOpen
  \bibinfo {note} {It should be noted that other methods adopting an
  oscillating basis such as Chebyshev polynomials may suffer from this
  problem.}\BibitemShut {Stop}%
\bibitem [{Note3()}]{Note3}%
  \BibitemOpen
  \bibinfo {note} {The Pad\'{e}\protect \xspace method is very fast, and so the
  time to estimate $\rho ^\protect \text {Pad\'{e}\protect \xspace }$ and
  $\sigma _i^\protect \text {Pad\'{e}\protect \xspace }$ is
  negligible.}\BibitemShut {Stop}%
\bibitem [{\citenamefont {Yoshimi}\ \emph {et~al.}(2019)\citenamefont
  {Yoshimi}, \citenamefont {Otsuki}, \citenamefont {Motoyama}, \citenamefont
  {Ohzeki},\ and\ \citenamefont {Shinaoka}}]{Yoshimi:2019bt}%
  \BibitemOpen
  \bibfield  {author} {\bibinfo {author} {\bibfnamefont {K.}~\bibnamefont
  {Yoshimi}}, \bibinfo {author} {\bibfnamefont {J.}~\bibnamefont {Otsuki}},
  \bibinfo {author} {\bibfnamefont {Y.}~\bibnamefont {Motoyama}}, \bibinfo
  {author} {\bibfnamefont {M.}~\bibnamefont {Ohzeki}},\ and\ \bibinfo {author}
  {\bibfnamefont {H.}~\bibnamefont {Shinaoka}},\ }\href
  {https://doi.org/10.1016/j.cpc.2019.07.001} {\bibfield  {journal} {\bibinfo
  {journal} {Computer Physics Communications}\ }\textbf {\bibinfo {volume}
  {244}},\ \bibinfo {pages} {319} (\bibinfo {year} {2019})}\BibitemShut
  {NoStop}%
\bibitem [{\citenamefont {Georges}\ \emph {et~al.}(1996)\citenamefont
  {Georges}, \citenamefont {Kotliar}, \citenamefont {Krauth},\ and\
  \citenamefont {Rozenberg}}]{RevModPhys.68.13}%
  \BibitemOpen
  \bibfield  {author} {\bibinfo {author} {\bibfnamefont {A.}~\bibnamefont
  {Georges}}, \bibinfo {author} {\bibfnamefont {G.}~\bibnamefont {Kotliar}},
  \bibinfo {author} {\bibfnamefont {W.}~\bibnamefont {Krauth}},\ and\ \bibinfo
  {author} {\bibfnamefont {M.~J.}\ \bibnamefont {Rozenberg}},\ }\href
  {https://doi.org/10.1103/RevModPhys.68.13} {\bibfield  {journal} {\bibinfo
  {journal} {Rev. Mod. Phys.}\ }\textbf {\bibinfo {volume} {68}},\ \bibinfo
  {pages} {13} (\bibinfo {year} {1996})}\BibitemShut {NoStop}%
\bibitem [{SpM()}]{SpM}%
  \BibitemOpen
  \href@noop {} {}\bibinfo {howpublished}
  {\url{https://spm-lab.github.io/SpM/manual/build/html/index.html}}\BibitemShut
  {NoStop}%
\bibitem [{\citenamefont {Shinaoka}\ \emph {et~al.}(2018)\citenamefont
  {Shinaoka}, \citenamefont {Otsuki}, \citenamefont {Haule}, \citenamefont
  {Wallerberger}, \citenamefont {Gull}, \citenamefont {Yoshimi},\ and\
  \citenamefont {Ohzeki}}]{PhysRevB.97.205111}%
  \BibitemOpen
  \bibfield  {author} {\bibinfo {author} {\bibfnamefont {H.}~\bibnamefont
  {Shinaoka}}, \bibinfo {author} {\bibfnamefont {J.}~\bibnamefont {Otsuki}},
  \bibinfo {author} {\bibfnamefont {K.}~\bibnamefont {Haule}}, \bibinfo
  {author} {\bibfnamefont {M.}~\bibnamefont {Wallerberger}}, \bibinfo {author}
  {\bibfnamefont {E.}~\bibnamefont {Gull}}, \bibinfo {author} {\bibfnamefont
  {K.}~\bibnamefont {Yoshimi}},\ and\ \bibinfo {author} {\bibfnamefont
  {M.}~\bibnamefont {Ohzeki}},\ }\href
  {https://doi.org/10.1103/PhysRevB.97.205111} {\bibfield  {journal} {\bibinfo
  {journal} {Phys. Rev. B}\ }\textbf {\bibinfo {volume} {97}},\ \bibinfo
  {pages} {205111} (\bibinfo {year} {2018})}\BibitemShut {NoStop}%
\bibitem [{\citenamefont {Wang}\ \emph {et~al.}(2020)\citenamefont {Wang},
  \citenamefont {Nomoto}, \citenamefont {Nomura}, \citenamefont {Shinaoka},
  \citenamefont {Otsuki}, \citenamefont {Koretsune},\ and\ \citenamefont
  {Arita}}]{PhysRevB.102.134503}%
  \BibitemOpen
  \bibfield  {author} {\bibinfo {author} {\bibfnamefont {T.}~\bibnamefont
  {Wang}}, \bibinfo {author} {\bibfnamefont {T.}~\bibnamefont {Nomoto}},
  \bibinfo {author} {\bibfnamefont {Y.}~\bibnamefont {Nomura}}, \bibinfo
  {author} {\bibfnamefont {H.}~\bibnamefont {Shinaoka}}, \bibinfo {author}
  {\bibfnamefont {J.}~\bibnamefont {Otsuki}}, \bibinfo {author} {\bibfnamefont
  {T.}~\bibnamefont {Koretsune}},\ and\ \bibinfo {author} {\bibfnamefont
  {R.}~\bibnamefont {Arita}},\ }\href
  {https://doi.org/10.1103/PhysRevB.102.134503} {\bibfield  {journal} {\bibinfo
   {journal} {Phys. Rev. B}\ }\textbf {\bibinfo {volume} {102}},\ \bibinfo
  {pages} {134503} (\bibinfo {year} {2020})}\BibitemShut {NoStop}%
\bibitem [{\citenamefont {Wallerberger}\ \emph {et~al.}(2021)\citenamefont
  {Wallerberger}, \citenamefont {Shinaoka},\ and\ \citenamefont
  {Kauch}}]{PhysRevResearch.3.033168}%
  \BibitemOpen
  \bibfield  {author} {\bibinfo {author} {\bibfnamefont {M.}~\bibnamefont
  {Wallerberger}}, \bibinfo {author} {\bibfnamefont {H.}~\bibnamefont
  {Shinaoka}},\ and\ \bibinfo {author} {\bibfnamefont {A.}~\bibnamefont
  {Kauch}},\ }\href {https://doi.org/10.1103/PhysRevResearch.3.033168}
  {\bibfield  {journal} {\bibinfo  {journal} {Phys. Rev. Research}\ }\textbf
  {\bibinfo {volume} {3}},\ \bibinfo {pages} {033168} (\bibinfo {year}
  {2021})}\BibitemShut {NoStop}%
\bibitem [{\citenamefont {Shinaoka}\ \emph {et~al.}(2021)\citenamefont
  {Shinaoka}, \citenamefont {Otsuki}, \citenamefont {Kawamura}, \citenamefont
  {Takemori},\ and\ \citenamefont {Yoshimi}}]{dcore}%
  \BibitemOpen
  \bibfield  {author} {\bibinfo {author} {\bibfnamefont {H.}~\bibnamefont
  {Shinaoka}}, \bibinfo {author} {\bibfnamefont {J.}~\bibnamefont {Otsuki}},
  \bibinfo {author} {\bibfnamefont {M.}~\bibnamefont {Kawamura}}, \bibinfo
  {author} {\bibfnamefont {N.}~\bibnamefont {Takemori}},\ and\ \bibinfo
  {author} {\bibfnamefont {K.}~\bibnamefont {Yoshimi}},\ }\href
  {https://doi.org/10.21468/SciPostPhys.10.5.117} {\bibfield  {journal}
  {\bibinfo  {journal} {SciPost Phys.}\ }\textbf {\bibinfo {volume} {10}},\
  \bibinfo {pages} {117} (\bibinfo {year} {2021})}\BibitemShut {NoStop}%
\bibitem [{dco()}]{dcore_doc}%
  \BibitemOpen
  \href@noop {} {}\bibinfo {howpublished} {\url{
  https://issp-center-dev.github.io/DCore/index.html }}\BibitemShut {NoStop}%
\bibitem [{\citenamefont {Parcollet}\ \emph {et~al.}(2015)\citenamefont
  {Parcollet}, \citenamefont {Ferrero}, \citenamefont {Ayral}, \citenamefont
  {Hafermann}, \citenamefont {Krivenko}, \citenamefont {Messio},\ and\
  \citenamefont {Seth}}]{triqs}%
  \BibitemOpen
  \bibfield  {author} {\bibinfo {author} {\bibfnamefont {O.}~\bibnamefont
  {Parcollet}}, \bibinfo {author} {\bibfnamefont {M.}~\bibnamefont {Ferrero}},
  \bibinfo {author} {\bibfnamefont {T.}~\bibnamefont {Ayral}}, \bibinfo
  {author} {\bibfnamefont {H.}~\bibnamefont {Hafermann}}, \bibinfo {author}
  {\bibfnamefont {I.}~\bibnamefont {Krivenko}}, \bibinfo {author}
  {\bibfnamefont {L.}~\bibnamefont {Messio}},\ and\ \bibinfo {author}
  {\bibfnamefont {P.}~\bibnamefont {Seth}},\ }\href
  {https://doi.org/http://dx.doi.org/10.1016/j.cpc.2015.04.023} {\bibfield
  {journal} {\bibinfo  {journal} {Computer Physics Communications}\ }\textbf
  {\bibinfo {volume} {196}},\ \bibinfo {pages} {398 } (\bibinfo {year}
  {2015})}\BibitemShut {NoStop}%
\bibitem [{\citenamefont {Werner}\ \emph {et~al.}(2006)\citenamefont {Werner},
  \citenamefont {Comanac}, \citenamefont {de' Medici}, \citenamefont {Troyer},\
  and\ \citenamefont {Millis}}]{Werner2006}%
  \BibitemOpen
  \bibfield  {author} {\bibinfo {author} {\bibfnamefont {P.}~\bibnamefont
  {Werner}}, \bibinfo {author} {\bibfnamefont {A.}~\bibnamefont {Comanac}},
  \bibinfo {author} {\bibfnamefont {L.}~\bibnamefont {de' Medici}}, \bibinfo
  {author} {\bibfnamefont {M.}~\bibnamefont {Troyer}},\ and\ \bibinfo {author}
  {\bibfnamefont {A.~J.}\ \bibnamefont {Millis}},\ }\href
  {https://doi.org/10.1103/physrevlett.97.076405} {\bibfield  {journal}
  {\bibinfo  {journal} {Physical Review Letters}\ }\textbf {\bibinfo {volume}
  {97}},\ \bibinfo {pages} {076405} (\bibinfo {year} {2006})}\BibitemShut
  {NoStop}%
\bibitem [{\citenamefont {Hafermann}\ \emph {et~al.}(2013)\citenamefont
  {Hafermann}, \citenamefont {Werner},\ and\ \citenamefont
  {Gull}}]{Hafermann2013}%
  \BibitemOpen
  \bibfield  {author} {\bibinfo {author} {\bibfnamefont {H.}~\bibnamefont
  {Hafermann}}, \bibinfo {author} {\bibfnamefont {P.}~\bibnamefont {Werner}},\
  and\ \bibinfo {author} {\bibfnamefont {E.}~\bibnamefont {Gull}},\ }\href
  {https://doi.org/10.1016/j.cpc.2012.12.013} {\bibfield  {journal} {\bibinfo
  {journal} {Computer Physics Communications}\ }\textbf {\bibinfo {volume}
  {184}},\ \bibinfo {pages} {1280} (\bibinfo {year} {2013})}\BibitemShut
  {NoStop}%
\bibitem [{ALP()}]{ALPS-CTHYB-SEGMENT}%
  \BibitemOpen
  \href@noop {} {}\bibinfo {howpublished} {\url{
  https://github.com/ALPSCore/CT-HYB-SEGMENT }}\BibitemShut {NoStop}%
\bibitem [{\citenamefont {Gaenko}\ \emph {et~al.}(2017)\citenamefont {Gaenko},
  \citenamefont {Antipov}, \citenamefont {Carcassi}, \citenamefont {Chen},
  \citenamefont {Chen}, \citenamefont {Dong}, \citenamefont {Gamper},
  \citenamefont {Gukelberger}, \citenamefont {Igarashi}, \citenamefont
  {Iskakov}, \citenamefont {K\"{o}nz}, \citenamefont {LeBlanc}, \citenamefont
  {Levy}, \citenamefont {Ma}, \citenamefont {Paki}, \citenamefont {Shinaoka},
  \citenamefont {Todo}, \citenamefont {Troyer},\ and\ \citenamefont
  {Gull}}]{ALPSCore2017}%
  \BibitemOpen
  \bibfield  {author} {\bibinfo {author} {\bibfnamefont {A.}~\bibnamefont
  {Gaenko}}, \bibinfo {author} {\bibfnamefont {A.}~\bibnamefont {Antipov}},
  \bibinfo {author} {\bibfnamefont {G.}~\bibnamefont {Carcassi}}, \bibinfo
  {author} {\bibfnamefont {T.}~\bibnamefont {Chen}}, \bibinfo {author}
  {\bibfnamefont {X.}~\bibnamefont {Chen}}, \bibinfo {author} {\bibfnamefont
  {Q.}~\bibnamefont {Dong}}, \bibinfo {author} {\bibfnamefont {L.}~\bibnamefont
  {Gamper}}, \bibinfo {author} {\bibfnamefont {J.}~\bibnamefont {Gukelberger}},
  \bibinfo {author} {\bibfnamefont {R.}~\bibnamefont {Igarashi}}, \bibinfo
  {author} {\bibfnamefont {S.}~\bibnamefont {Iskakov}}, \bibinfo {author}
  {\bibfnamefont {M.}~\bibnamefont {K\"{o}nz}}, \bibinfo {author}
  {\bibfnamefont {J.}~\bibnamefont {LeBlanc}}, \bibinfo {author} {\bibfnamefont
  {R.}~\bibnamefont {Levy}}, \bibinfo {author} {\bibfnamefont {P.}~\bibnamefont
  {Ma}}, \bibinfo {author} {\bibfnamefont {J.}~\bibnamefont {Paki}}, \bibinfo
  {author} {\bibfnamefont {H.}~\bibnamefont {Shinaoka}}, \bibinfo {author}
  {\bibfnamefont {S.}~\bibnamefont {Todo}}, \bibinfo {author} {\bibfnamefont
  {M.}~\bibnamefont {Troyer}},\ and\ \bibinfo {author} {\bibfnamefont
  {E.}~\bibnamefont {Gull}},\ }\href
  {https://doi.org/10.1016/j.cpc.2016.12.009} {\bibfield  {journal} {\bibinfo
  {journal} {Computer Physics Communications}\ }\textbf {\bibinfo {volume}
  {213}},\ \bibinfo {pages} {235} (\bibinfo {year} {2017})}\BibitemShut
  {NoStop}%
\bibitem [{\citenamefont {Wallerberger}\ \emph {et~al.}(2018)\citenamefont
  {Wallerberger}, \citenamefont {Iskakov}, \citenamefont {Gaenko},
  \citenamefont {Kleinhenz}, \citenamefont {Krivenko}, \citenamefont {Levy},
  \citenamefont {Li}, \citenamefont {Shinaoka}, \citenamefont {Todo},
  \citenamefont {Chen}, \citenamefont {Chen}, \citenamefont {LeBlanc},
  \citenamefont {Paki}, \citenamefont {Terletska}, \citenamefont {Troyer},\
  and\ \citenamefont {Gull}}]{ALPSCore2018}%
  \BibitemOpen
  \bibfield  {author} {\bibinfo {author} {\bibfnamefont {M.}~\bibnamefont
  {Wallerberger}}, \bibinfo {author} {\bibfnamefont {S.}~\bibnamefont
  {Iskakov}}, \bibinfo {author} {\bibfnamefont {A.}~\bibnamefont {Gaenko}},
  \bibinfo {author} {\bibfnamefont {J.}~\bibnamefont {Kleinhenz}}, \bibinfo
  {author} {\bibfnamefont {I.}~\bibnamefont {Krivenko}}, \bibinfo {author}
  {\bibfnamefont {R.}~\bibnamefont {Levy}}, \bibinfo {author} {\bibfnamefont
  {J.}~\bibnamefont {Li}}, \bibinfo {author} {\bibfnamefont {H.}~\bibnamefont
  {Shinaoka}}, \bibinfo {author} {\bibfnamefont {S.}~\bibnamefont {Todo}},
  \bibinfo {author} {\bibfnamefont {T.}~\bibnamefont {Chen}}, \bibinfo {author}
  {\bibfnamefont {X.}~\bibnamefont {Chen}}, \bibinfo {author} {\bibfnamefont
  {J.~P.~F.}\ \bibnamefont {LeBlanc}}, \bibinfo {author} {\bibfnamefont
  {J.~E.}\ \bibnamefont {Paki}}, \bibinfo {author} {\bibfnamefont
  {H.}~\bibnamefont {Terletska}}, \bibinfo {author} {\bibfnamefont
  {M.}~\bibnamefont {Troyer}},\ and\ \bibinfo {author} {\bibfnamefont
  {E.}~\bibnamefont {Gull}},\ }\href@noop {} {\bibinfo {title} {Updated core
  libraries of the {ALPS} project}} (\bibinfo {year} {2018}),\ \Eprint
  {https://arxiv.org/abs/arXiv:1811.08331} {arXiv:1811.08331} \BibitemShut
  {NoStop}%
\bibitem [{Note4()}]{Note4}%
  \BibitemOpen
  \bibinfo {note} {ALPS/CTHYB-segment has a parameter for specifying the
  maximum runtime of code, which is set as 60 second in this
  demonstration.}\BibitemShut {Stop}%
\end{thebibliography}
%

\end{document}